\documentstyle[12pt]{article}

\newtheorem{satz}{Satz}[section]
\newtheorem{lemma}[satz]{Lemma}
\newtheorem{definition}{Definition}[section]
\newtheorem{behauptung}[satz]{Behauptung}
\newcommand{\BS}{\begin{satz}}
\newcommand{\ES}{\end{satz}}
\newcommand{\BDEF}{\begin{definition}}
\newcommand{\EDEF}{\end{definition}}
\newcommand{\BB}{\begin{behauptung}}
\newcommand{\EB}{\end{behauptung}}
\newcommand{\BL}{\begin{lemma}}
\newcommand{\EL}{\end{lemma}}
\newcommand{\BD}{\begin{displaymath}}
\newcommand{\ED}{\end{displaymath}}
\newcommand{\BE}{\begin{equation}}
\newcommand{\EE}{\end{equation}}
\newcommand{\BEA}{\begin{eqnarray}}
\newcommand{\EEA}{\end{eqnarray}}
\newcommand{\BEAS}{\begin{eqnarray*}}
\newcommand{\EEAS}{\end{eqnarray*}}
\newcommand{\BA}{\begin{array}}
\newcommand{\EA}{\end{array}}
\newcommand{\w}{\wedge}
\newcommand{\NN}{\nonumber}
\newcommand{\NI}{\noindent}
\newcommand{\IN}{\indent}
\newcommand{\G}{\left}
\newcommand{\D}{\right}
\newcommand{\GA}{\langle}
\newcommand{\DA}{\rangle}
\newcommand{\DS}{\displaystyle}
\newcommand{\TS}{\textstyle}

\newcommand{\SSS}{\scriptscriptstyle}
\newcommand{\I}{\int_\Sigma}

\newcommand{\V}[1]{{\vec #1}}
\newcommand{\B}[1]{{\bar #1}}

\renewcommand{\P}{\partial}

\newcommand{\F}{\frac}
\newcommand{\comment}[1]{}
\newcommand{\nummer}[1]{
  \hskip #1
  \refstepcounter{equation}\rlap{\@eqnnum }
  \hskip -#1 }
\newcommand{\setR}{\ifmmode{I\hskip -4pt R}
    \else{\hbox{$I\hskip -4pt R$}}\fi} 
\newcommand{\setQ}{\ifmmode{Q\hskip-5.0pt\vrule height6.0pt depth 0pt\hskip6pt}
    \else{\hbox{$Q\hskip -5.0pt\vrule height6.0pt depth 0pt\hskip6pt$}}\fi}
\newcommand{\setZ}{\ifmmode{Z\hskip -6pt /}
    \else{\hbox{$Z\hskip -6pt /$}}\fi}
\newcommand{\St}[1]{|_{{}_{{}_{#1}}}}

\renewcommand{\phi}{\varphi}

\newcommand{\aut}{\mbox{aut}}
\newcommand{\Aut}{\mbox{Aut}}
\newcommand{\ad}{\mbox{ad}}
\newcommand{\Ad}{\mbox{Ad}}
\newcommand{\Tr}{\mbox{tr\hspace{0.5ex}}}
\newcommand{\Autg}{\mbox{Aut\BBB }}
\newcommand{\autg}{\mbox{aut\BBB }}
\newcommand{\lra}{\longrightarrow}
\newcommand{\Lra}{\Longrightarrow}
\newcommand{\lmt}{\longmapsto}

\newcommand{\Llra}{\Longleftrightarrow}
\newfont{\euler}{eufm10 scaled \magstephalf}
\newcommand{\BBB}{{\euler g}}
\newcommand{\X}{{\euler X}}

\renewcommand{\S}{\Sigma}

\newcommand{\pr}{\NI{\bf Proof:}}

\begin{document}
\begin{center}
{\Large  Hamiltonian reduction of Bianchi Cosmologies}\\[1cm]
{\large Joachim Schirmer}\\[5mm]
Fakult\"at f\"ur Physik der Universit\"at Freiburg\\
Hermann-Herder-Str. 3, 79104 Freiburg i.Br. / FRG
\end{center}

\begin{abstract}

It was noted recently that the ADM-diffeomorphism-constraint does not
generate all observed symmetries for several Bianchi-models. We will suggest
not to use the ADM-constraint restricted to homogeneous variables, but some
equivalent which is derived from a restricted action principle. This will
generate all homogeneity preserving diffeomorphisms, which will be
shown to be automorphism generating vector fields, in class A and class B
models. Following Dirac's constraint formalism one will naturally be
restricted to the unimodular part of the automorphism group.
\end{abstract}

\hfill\begin{minipage}{5cm}
\begin{center}
University of Freiburg\\
July 1994\\
THEP 94/21
\end{center}
\end{minipage}
\vspace{2cm}

\section{3-dimensional Lie-groups}
In this preliminary paragraph we quickly recapitulate the well known
Bianchi-Behr classification of 3-dimensional Lie algebras \cite{JI,Behr}. Up
to isomorphy there exists exactly one simply-connected Lie-group associated
with each type of algebra. The structure constant tensor is represented
by a symmetric tensor-density and a covector:
\BEA
 &C^a{}_{bc}=
  n^{ad}\epsilon_{bcd}+a_f\delta^{fa}_{bc}&\\
\label{5}
 &n^{ad}=\epsilon^{bc(d}C^{a)}{}_{bc}\qquad a_f={\TS\F{1}{2}}C^a{}_{fa}&
\EEA
The Jacobi-identity yields:
\BE
\label{1}
   a_cC^c{}_{ab}=0=a_cn^{cd}
\EE
Transforming the basis of the Lie algebra by $b_a\mapsto\B
b_{a'}=b_a{A^{-1}}^a{}_{a'}$ the structure-coefficients change
as follows:
\BD
  \BA{rcccl}
  C^a{}_{bc}&\longmapsto&
   \B C^{a'}{}_{b'c'}&=&A^{a'}{}_aC^a{}_{bc}{A^{-1}}^b{}_{b'}
          {A^{-1}}^c{}_{c'}\\[0.8ex]
  a_b&\longmapsto&
    \B a_{b'}&=&a_b{A^{-1}}^b{}_{b'}\\[0.8ex]
  n^{ab}&\longmapsto&\B n^{a'b'}&=&{\TS\F{1}{\det A}}n^{ab}A^{a'}{}_a
  A^{b'}{}_b
  \EA
\ED
Choosing a suitable basis we can diagonalize $n$ such that
$n=$diag$(n^1,n^2,n^3)$ and assume that $a_a=(0,0,a)$. One distinguishes
algebras of class A with $a=0$ and B where $a\not=0$. For class B algebras
rg$\,n\leq 2 $ holds because of (\ref{1}). In the special case of rg$\,n=2,
{}~a\not=0,~a$ cannot be normalized because of the determinant in the
transformation law for n, so that these classes are 1-parameter-classes.
\bigskip

\centerline{\begin{tabular}{cc|cccccc}
cl  & Bianchi-type & $a$ & $n^1$ & $n^2$ & $n^3$ & dim Aut\BBB & dim Int\BBB\\
\hline
A   &   I         &  0  &   0   &   0   &   0   &      9      &      0     \\
    &   II        &  0  &   1   &   0   &   0   &      6      &      2     \\
    &   VI$_0$    &  0  &   1   &  -1   &   0   &      4      &      3     \\
    &   VII$_0$   &  0  &   1   &   1   &   0   &      4      &      3     \\
    &   VIII      &  0  &   1   &   1   &  -1   &      3      &      3     \\
    &   IX        &  0  &   1   &   1   &   1   &      3      &      3     \\
\hline
B   &   V         &  1  &   0   &   0   &   0   &      6      &      3     \\
    &   IV        &  1  &   1   &   0   &   0   &      4      &      3     \\
    &   III=VI$_1$&  1  &   1   &  -1   &   0   &      4      &      2     \\
    &   VI$_a$ 	  &  a  &   1   &  -1   &   0   &      4      &      3     \\
    &   VII$_a$   &  a  &   1   &   1   &   0   &      4      &      3     \\
\end{tabular}}
Considering the Lie algebras we can conclude that the associated groups of
B-class-algebras are not compact. We recapitulate the proof given in
\cite{A&S}. Let $\chi_i=\F{1}{2}\epsilon_{ijk}b^j\w b^k$ be three linearly
independent left invariant 2-forms. Assuming that $G$ is compact Stoke's
theorem leads to a contradiction:
\BEAS
 0&=&\int_{\P G}\chi_i=\int_Gd\chi_i=-\int_G\epsilon_{ijk}
     C^j{}_{lm}b^l\w b^m\w b^k\\
  &=&-2C^j{}_{ij} \mbox{vol}_b(G)=-4a_i\mbox{vol}_b(G)\quad\Lra\quad a_i=0
\EEAS

The dimension of the automorphism group of the Lie algebra is in all cases
easily determined by considering the isotropy subgroup of
the action of $Gl(3,\setR)$ on the structure constants. Every group
automorphism induces an automorphism of the Lie algebra \BBB\ $=T_eG$, but
only if $G$ is simply-connected there is a 1:1-mapping between group and
algebra automorphisms \cite{Warner}. The group of algebra automorphisms as a
group of linear endomorphisms is easily investigated, whereas the group of
group automorphisms is only known to be generally smaller unless in the
simply-connected case. We will now give the characterizing property of
automorphism vector fields, i.e. vector fields the associated flow of which
consists of automorphisms.

\NI{\bf Proposition:}
\BD
Z\in\aut G \qquad\mbox{if and only if}\qquad Z(p\cdot
q)=TL_pZ(q)+TR_qZ(p)\qquad
\forall p,q\in G.
\ED
\pr
\BEAS
   &Z\in\aut G\quad \Llra\quad\Phi_t^Z\in\Aut G\quad\Llra
   \quad\Phi_t^Z(pq)=\Phi_t^Z(p)\Phi_t^Z(q) \quad\Llra&\\
   &Z(\Phi_t^Z(pq))=TL_{\Phi_t^Z(p)}Z(\Phi_t^Z(q)+TR_{\Phi_t^Z(q)}\Phi_t^Z(p)
\EEAS
Whereas there is only a 1:1-mapping between group and algebra automorphisms
in the simply-connected case there exists a 1:1-mapping always for the case
of inner automorphisms: Int\BBB\ $\approx G/C\approx\mbox{Int}G$, where $C$
is the center of the group. So it is natural to ask for the
automorphism vector field ($\in$ int$G$) given by the element
$X\in$ \BBB\  of the Lie algebra, which directly defines $\ad_X\in$ int\BBB.

\NI{\bf Proposition:}\\
The inner derivation $\ad_X$ associated to the element $X\in T_eG$ of the Lie
algebra defines a vector field of int$G$ given by
\BE
   Z(p)=X^R(p)-X^L(p):=TR_pX-TL_pX
\EE
\pr\\
$Z$ generates inner automorphisms since right and left invariant vector
fields
commute and the left multiplication, regarded as an operation of the group on
itself, has right invariant fundamental vector fields and vice versa:
\BEAS
 &\Phi_t^Z=\lim\limits
  _{n\to\infty}\G(\Phi_{t/n}^{X^R}\circ\Phi_{t/n}^{X^L}\D)^n=
    \Phi_t^{X^R}\Phi_{-t}^{X^L}=L_{\exp tX}R_{\exp -tX}&\\
 &T\Phi_t^Z=T\, L_{\exp tX}R_{\exp -tX} = \Ad_{\exp tX} = e^{t\ad_X}&
\EEAS
For the convenience of the reader we note that the Christoffel-symbols of the
Levi-Civita-connection with respect to a left invariant basis
$\{b_b\}$ expressed by structure constants take the form:
\BE
\label{2}
  \Gamma^a{}_{bc}={\TS\F{1}{2}}(C^a{}_{bc}+C_b{}^a{}_c+C_c{}^a{}_b)
\EE
The connection form is as usual
\BE
  \omega^a{}_b=\Gamma^a{}_{cb}b^c\qquad.
\EE

\section{Kinematical Preliminaries}
In this paragraph we will set the kinematical stage. We will carefully
analyze the question of what the admissible shift vector fields are.

\NI{\bf Definition:} A homogeneous cosmological model is a
Lorentz-manifold $(M,g)$ on which a 3-dimensional isometry group acts simply
transitively on spatial hypersurfaces.

This definition is chosen for simplicity. One could imagine to define more
generally that an isometry group acts forming 3-dimensional spacelike orbits.
In this case these orbits are homogeneous Riemannian spaces. One can classify
the simply-connected 3-dimensional homogeneous Riemannian spaces according to
the dimension of the isometry group and it turns out that apart from the
Kantowski-Sachs-case there is a three-dimensional subgroup acting
simply-transitive on the orbits \cite{Collins}. Discarding the
Kantowski-Sachs-case the simply-connected models in this more general
definition are contained in our definition.

A homogeneous cosmological model is (time-locally) isometric to $(\setR\times
G, -ds\otimes ds + q_{ab}(s)a^a\otimes a^b)$, where $\{a^a\}_{a=1,2,3}$ is a
left invariant basis of $T^*(G)$. We sketch the proof
\cite{MacCallum,Straumann}:\\
\IN Consider the normalized geodesic at a point $p$, normal to the orbit of
$G$. The orbit at the point $q$ reached by the geodesic at a certain
parameter value is parallel to the original orbit, i.e. taking the
normalized geodesic at another point $p'$ of the original orbit in normal
direction one reaches the orbit of $q$ at the same parameter value.

The metric $(-ds\otimes ds + q_{ab}a^a\otimes a^b)$ is often used as the
starting point to evaluate Einstein's equations for homogeneous models. We
notice that our space-time is naturally foliated in space and time and that
our metric is represented in a lapse-1-shift-0-representation. For
conceptional reasons it is necessary to check if this is choice of lapse and
shift is the only possible, in which the model is described by a
time-dependent left invariant metric. In the normal ADM-framework the
secondary constraints associated with lapse and shift reflect the gauge
freedom and are equivalent to the equations $G(n,n)=0$ and $G(n,a_a)=0$.
If our model required a certain gauge these constraints are satisfied
identically. The question of possible shift vector fields is also of
practical importance since a suitable choice might simplify the matrix
$q_{ab}$ and help in the search of solutions. A shift vector field generates
a diffeomorphism of the foliating spacelike surface. If this diffeomorphism
shall not destroy spatial homogeneity it has to leave left invariant metrics
left invariant.

\NI{\bf Lemma}: If a diffeomorphism $\psi$ maps every left invariant metric
of
a connected Lie-group onto a left invariant metric, it has the form
$\psi=L_{\psi(e)}\circ\phi$, where $\phi$ is an automorphism.

\pr \\
Let $\phi:=L_{\psi(e)^{-1}}\circ\psi$ and $\xi^{\SSS L},~\eta^{\SSS
L}$ the left invariant vector fields associated with the vectors
$\xi,~\eta\in T_eG$ and $q$ a left invariant metric.
\BEAS
  \phi^*q\;(g)\G(\xi^{\SSS L}(g),\eta^{\SSS L}(g)\D)&=&
       q\G(\phi(g)\D)\Big(T\phi\;TL_g\xi,\;T\phi\;TL_g\eta\Big)\\
   &=&q(e)\Big(TL_{\phi(g)}^{-1}\;T\phi\;TL_g\xi,\;TL_{\phi(g)}^{-1}\;T\phi\;
   TL_g\eta\Big)\\
  \phi^*q\;(g)\G(\xi^{\SSS L}(g),\eta^{\SSS L}(g)\D)&=&
       \phi^*q(e)(\xi,\eta)=q(e)(T\phi\xi,T\phi\eta)
\EEAS
Since this holds for arbitrary metrics, one concludes
\BD
   T_g\phi T_eL_g=T_eL_{\phi(g)}T_e\phi\qquad\forall g\in G
\ED
and hence $(T_e\phi\xi)^{\SSS L}(\phi(g))=T_g\phi(\xi^{\SSS L}(g))$. Then
$\phi(g\exp t\xi)$ and $\phi(g)\phi(\exp t\xi)$ are integral curves of the
vector field $(T_e\phi\xi)^{\SSS L}$ and thus they agree. Since a connected
group is generated by a neighbourhood of the unit element, $\phi$ is an
automorphism and the lemma is proved.

Now we can write down the most general time vector field preserving spatial
homogeneity.
\BE
  \F{\P}{\P t}=N(t)n+\V N=N(t)\F{\P}{\P s}+ \V N\qquad \V N\in\mbox{aut}G
\EE
The lapse-function can only depend on time, since the normal
vector field is a natural vector field $n=\F{\P}{\P s}$. The dual relation of
$s$ and $t$ is $n^\flat=ds=Ndt$ and application of the exterior derivative
on both sides shows that the lapse function must not depend on
space-coordinates. The interpretation is simple: If the lapse-function also
depended on space-coordinates the slices of simultaneity would not coincide
with the group orbits. The admissible shift vector field must be a
sum of a right\-invariant vector field and an element of $\aut$\BBB, the
Lie algebra of the group of automorphisms of $G$. The right invariant
vector field generates only lefttranslations which leave the metric unchanged.
So it has a trivial action on the metric and will be disregarded.
For Bianchi-type VIII and IX all automorphisms are inner. According to our
proposition of the preceding paragraph a vector field generating inner
automorphisms has the form $\V N=\xi^{\SSS R}-\xi^{\SSS L}$ for a certain
$\xi\in$ \BBB\ $=T_eG$. Neglecting the right invariant field one can choose a
left invariant shiftvector field $\V N\in$ \BBB\ = \X$^L=\{\mbox{leftinv.
vector fields}\}$ without loss of generality. But this is only correct for
these
two Bianchi-types whereas in all other cases the automorphism group is larger
than the inner automorphism group. For the Bianchi-type I, II and III the
associated groups even have a nontrivial center. Thus we stick to the
more general assumption that the shift is an element of aut$G$ and note that
therefore the coefficients of these fields with respect to a left invariant
basis need not to be constant. This must be taken into account when one asks
for the degrees of freedom in the reduced phase space.

We first consider the Bianchi-IX-case. There exists a
simply-connected compact group, $SU(2)$, with group manifold $S^3$. There is
a 1:1-relation between group- and algebra-automorphisms and all automorphisms
are inner automorphisms. Even in the non-simply-connected case of $SO(3)$
with underlying manifold $\setR P^3$ every automorphism of the Lie algebra
defines an automorphism of the group, since every automorphism of $SU(2)$
induces an automorphism of $SO(3)$. In all other Bianchi-types the
universal covering group is not compact. In order to obtain a
compact group one must divide by a compact normal subgroup. This
procedure reduces the automorphism group since identified points must be
mapped onto identified points. The effect can easily be seen in the
Bianchi-I-case. The universal covering group is $(\setR^3,+)$, the
automorphism group is $Gl(3,\setR)$. After compactification to $(T^3,+)$ only
a discrete subgroup of $Gl(3,\setR)$ is left, $SL(3,Z)$. This means that
we cannot introduce a shift vector field without destroying manifest
homogeneity, the gauge is fixed. Of course it is still possible to use a
right invariant shift, which has no effect to the metric.

The consequences for the Hamiltonian description can be anticipated. In the
case of the torus $(T^3,+)$ one can use the normal
ADM-diffeomorphism-constraint, since integration by parts is possible for
this compact model. But after restriction to homogeneous
metrics and the evaluation of the covariant derivative in terms of structure
constants the constraint vanishes identically, so it does not generate any
gauge. In the case of $(\setR,+)$ one should not use the ADM-constraint, but
some analogon which is found by restricting the action principle to
left invariant metrics and avoiding integration by parts. This analogon
generates all automorphisms. For the $SU(2)$-model both descriptions
coincide.

After introducing a shift vector field it is necessary to change the
left invariant basis, since it is no longer time-independent $\F{d}{dt}a_a=
[\F{\P}{\P t},a_a]=0$, whereas one had $[\F{\P}{\P s},a_a]=0$. By applying a
time-dependent Lie-algebra-automorphism $b_b:=a_a{S^{-1}}^a{}_b(t),\quad
S:\setR\lra\Autg$, the condition $[\F{\P}{\P t},b_b]=0$ which is needed for
the ADM-formalism can be satisfied:
\BEAS
  b_b&=&a_a{{S^{-1}}}^a{}_b\qquad\qquad S:\setR\longrightarrow \Autg\\
   0=[\F{\P}{\P t},b_b]&=&a_a{\dot{S^{-1}}}^a{}_b
         +[N\F{\P}{\P s}+\V N,a_a]{S^{-1}}^a{}_b\\
     &=&b_cS^c{}_a\dot{S^{-1}}^a{}_b+L_\V Nb_b=0\\
  \Llra\qquad    \dot{S}^c{}_a{S^{-1}}^a{}_b&=&b^c(L_\V N
                      b_b)=:
          {A(\V N)}^c{}_b
          \qquad A:\setR\longrightarrow\autg
\EEAS
Substitution of the basis $\{ds,a^a\}$ by $\{dt,b^b\}$ leads to a change of
the metric:
\BEAS
  ds&=&Ndt\hspace{10ex} a^a~=~{S^{-1}}^a{}_b(N^bdt+b^b)\hspace{8ex}
     N^b:=b^b(\V N)\\
   g&=&-N^2dt\otimes dt+{S^{-1}}^c{}_aq_{cd}{S^{-1}}^d{}_b(N^adt+b^a)\otimes
       (N^bdt+b^b)
\EEAS
The infinitesimal automorphism is generally not constant because the
automorphism field $A\in\autg$ may be time-dependent. It was suggested in
\cite{Henneaux} that it has to be constant and because of a localizability
requirement only inner automorphism are acceptable.

We can read-off the effect of the automorphism vector field to the spatial
metric components:
\BE
    q_{ab}(t)\lmt {S^{-1}}^c{}_a(t)q_{cd}(t){S^{-1}}^d{}_b(t)
\EE
Using the well known equation for the extrinsic curvature one can easily
check that the extrinsic curvature remains left invariant after introduction
of the shift vector field so that the dynamics of the homogeneous models can
be
regarded as a finite-dimensional mechanical system.
\BEA
        \dot{q}&=&2NK+L_\V Nq\\
\NN    K_{ab}b^a\otimes b^b&=&\F{1}{2N}\G[\dot{q}_{ab}b^a\otimes b^b
      -L_\V N\G(q_{ab}b^a\otimes b^b\D)\D]\\
  &=&\F{1}{2N}\G[\dot{q}_{ab}+q_{ac}{A(\V N)}^c{}_b+{A(\V N)}^c{}_aq_{cb}\D]
       b^a\otimes b^b
\EEA

\section{Hamiltonian formulation of homogeneous models}
Usually one specializes the ADM-constraint and the equations of motion to
left invariant metrics in order to gain a Hamiltonian description of
homogeneous models. But this "restriction procedure" is normally not allowed
since left invariant metrics and automorphism shifts are often not contained
in the mathematical spaces for which the ADM-framework was derived. This is
the reason why the usual diffeomorphism-constraint fails to generate all
automorphisms. Instead we will restrict the 3:1-split action-principle to
left invariant metrics, automorphism-shifts and spatially constant
lapse-functions. Then we can directly derive equivalents of the constraints
and as well equations of motion. According to a mathematical argument by
Palais \cite{Palais} both descriptions coincide in the case of compact
groups, but for class B algebras there are no compact groups and for the
compact Bianchi-I-model there are no nontrivial shift vector fields.

Starting from the 3:1-split action-principle we recapitulate the
derivation of the ADM-Constraints. Let
\BEA
 & L(q,\dot{q},N,\V N)={\DS\int_\Sigma}\eta N(K_{ab}K^{ab}-K^2+R)&\\
 & K_{ab}={\DS\F{1}{2N}}\G(\dot{q}_{ab}-L_\V Nq_{ab}\D)&
\EEA
where $q$ is the spatial 3-metric, $K$ the extrinsic curvature, $R$ the
Ricci-scalar of the spatial 3-metric, $N$ lapse and $N$ the
shift vector field.
\BEA
 &p^{ab}={\DS\F{\delta L}{\delta\dot{q}_{ab}}}=\G(K^{ab}-Kq^{ab}\D)\eta&\\
 &K^{ab}=*p^{ab}-{\TS\F{1}{2}}pq^{ab}&\\
 &p:=*p^{ab}q_{ab}=-2K&
\EEA
$p^{ab}$ is the momentum, a tensor valued 3-form. We can perform the
Legendre-transformation and obtain the Hamiltonian:
\BE
\label{3}
 H=\I \dot{q}_{ab}p^{ab}-L=\I p^{ab}L_\V Nq_{ab}+\I\eta N\G[\GA
 p^{ab}|p_{ab}\DA-{\TS\F{1}{2}}p^2-R\D]
\EE
Here $\GA|\DA$ denotes the metric product of 3-forms. Since the
time-derivatives of lapse and shift do not appear in the Lagrangian one
obtains primary constraints
\BE
   p_N=\F{\delta L}{\delta\dot{N}}\approx 0\qquad p_\V N=\F{\delta
   L}{\delta\dot{\V N}}\approx 0
\EE
and subsequently secondary constraints (which are equivalent to the
Euler-\linebreak Lagrange-equations associated to lapse and shift).
\BEA
   C
   _H&=&\{H,p_N\}=\eta\G(\GA p^{ab}|p_{ab}\DA-{\TS\F{1}{2}}p^2-R\D)\\
   {C_D}_c&=&\{H,p_{N^c}\}=-2q_{ac}Di_bp^{ab}\qquad,
\EEA
where $D$ is the exterior covariant derivative. These secondary constraints
must be regarded as functionals on a certain space of lapse functions or
shift vector fields. Especially in the case of the diffeomorphism-constraint
a
boundary integral was neglected:
\BEAS
  {C_D}_c&=&\{H,p_{N^c}\}=\F{\delta}{\delta N^c}\I p^{ab}L_\V Nq_{ab}\\
  \I p^{ab}L_\V Nq_{ab}&=&2\I p^{ab}N_{a;b}=2\I p^{ab}i_bDN_a\\
     &=&2\I i_bp^{ab}\w DN_a=2\int_{\P\S}N_ai_bp^{ab}-2\I N^c q_{ca}Di_bp^{ab}
\EEAS
So $H$ is only functionally differentiable if the boundary term vanishes and
this is a restriction to the space of admissible shift vector fields.

It is useful to remember the link between the 3:1-split Hamiltonian and
the space-time-description. Having chosen a foliation there is an adapted
tetrad $\{n,b_a\}_{a=1,2,3}$, where $n$ is normal, $b_a$ parallel to the
spacelike hypersurfaces. Then the Hamiltonian constraint $C_H=0$ is
equivalent to the Einstein-equation $G(n,n)=0$, the diffeomorphism constraint
${C_D}_c=0$ reflects the equation $G(n,b_a)=0$ and the equations of motion
represent the equations $G(b_a,b_b)=0$ .

When one restricts this ADM-framework to left invariant metrics, momentum
forms, automorphism-shifts and spatially constant lapse functions, and
substitutes $\S$ by $G$ and uses the formula (\ref{2})  for the
Levi-Civita-connection with respect to a left invariant basis one obtains in
case of the diffeomorphism constraint:
\BEA
\NN
  {C_D}_c\St{li}&=&-2q_{ab}Di_bp^{ab}=-2(q_{ac}di_bp^{ab}+q_{ac}\omega^a{}_d\w
     i_bp^{db})\\
\label{6}
  &=&-2q_{ab}C^a{}_{cd}p^{db}+4q_{ac}p^{ab}a_b=-2\Tr qk_cp+4(\V apq)_c
\EEA
where $a$ is as in (\ref{5}) and thus the second term vanishes for class A
algebras. If this restricted constraint is a useful analogon in the
homogeneous case it should generate all automorphisms of the
group. Since in the homogeneous case we deal with a finite-dimensional
mechanical system we can easily determine the momentum mapping for
the action of the automorphism group and compare it with this constraint. Our
configuration space is the space of all positive definite symmetric
(0,2)-tensors over the Lie algebra $\BBB=T_eG$. The tangent bundle is the
space of all symmetric (0,2)-tensors over $\BBB$, the cotangent-bundle
$T^*Q$ is the space of all symmetric (2,0)-tensors over the basis $Q$ and
the natural product is given by contraction. Using a basis $b_a\in T_eG$ one
identifies $TQ$ and $T^*Q$ with symmetric matrices over the base manifold of
all positive definite matrices, and the contraction is given by the
trace of the matrices' product. By the
choice of the basis the automorphism group of \BBB\ is identified
with a subgroup of $Gl(3,\setR)$. We now construct the momentum mapping of
the lifted group action on $T^*Q$ \cite{A&M}.
\BEAS
  &S\in\mbox{ Aut\BBB}\qquad q\in Q\quad \pi\in T^*Q \qquad&\\
  & \Phi:\Autg\times Q \longrightarrow Q \qquad
    \Phi^{T^*}:\Autg\times T^*Q \lra T^*Q &
\EEAS
\vspace{-20pt}
\BEA
\NN   \Phi(S,q)&=&{S^{-1}}^T q S^{-1}\\
\NN   \Phi^{T^*}(S,\pi)&=&S\pi S^T\\
      P(A)(\pi_q)=\pi\G(\F{d}{dt}\St{t=0}\Phi(e^{tA},q)\D)&=&\pi(-A^Tq-qA)
      =-2\Tr qA\pi\qquad
\EEA
Introducing a basis of the automorphism group of the Lie-algebra we can
easily compare the diffeomorphism constraint and the momentum mapping:\\[1ex]
\IN $\{r_\mu\}_{\mu=0,...,\dim\mbox{\footnotesize aut\BBB}-1}$ \qquad basis
    of aut\BBB \\
\IN $\{r_i\}_{i=1,...,\dim\mbox{\footnotesize aut\BBB}-1}$ \qquad basis
    saut\BBB\ $:=\{A\in\autg~ |~\Tr A=0\}$
\BE
   P(A)=-2\Tr qA\pi=-2A^\mu\Tr qr_\mu\pi
\EE
So the diffeomorphism-constraint (\ref{6}) contains two different problems:
\begin{enumerate}
\item
The second summand which is only present for class B algebras
has no meaning.
\item
Discarding the second summand -- restricting to class A algebras --
the diffeo\-morphism-constraint seems to generate only inner automorphisms.
\end{enumerate}
After the discussion of the preceding paragraph we are able to guess the
reasons of these shortcomings:
\begin{enumerate}
\item
Class B Lie groups are not compact. The boundary integral which appears in
the derivation of the constraint has to be defined as a limit over compact
subsets for which Stoke's theorem is valid), and this sequence does not tend
to zero, since an automorphism field cannot satisfy any asymptotic fall-off
conditions.
\item
Automorphism fields which do not generate inner automorphisms cannot have
constant components with respect to a left invariant basis. In order to
generate outer automorphisms it is necessary to integrate the constraint with
the appropriate shift vector field. Thus the restricted constraint is not
suitable for the finite dimensional model.
\end{enumerate}
If one instead derives the constraints from an action principle that is
restricted to left invariant metrics, they generate all automorphisms
regardless of class A or B algebras. Unfortunately the action integral is not
defined in the case of non-compact groups, but since the variational
principle holds locally we can integrate over a nonempty open subset $U$ with
compact closure.
\BEA
     &L(q,\dot{q},N,A)={\DS\int_U} N(K_{ab}K^{ab}-K^a{}_a^2+R)\eta
       =NV(K_{ab}K^{ab}-K^a{}_a^2+R)&\quad\\
\NN  &V:={\DS\int_U}\eta\qquad N=N(t)\qquad \V N\in\aut G&\\
\NN  &K={\DS\F{1}{2N}}(\dot{q}-L_\V N q)=\F{1}{2N}(\dot{q}+A^Tq+qA)&
\EEA
The momentum is :
\BE  \B p^{ab}=V(K^{ab}-K^c{}_cq^{ab})=\int_U p^{ab}\qquad
          K^{ab}=\F{1}{V}(\B p^{ab}-\B p^c{}_cq^{ab})
\EE
Performing the Legendre transformation one obtains
\BE
   H=-2A^\mu\Tr q r_\mu\B p
       +N\G[\F{1}{V}\G(\Tr (\B pq)^2-{\TS\F{1}{2}}\Tr\!^2\B pq\D)-VR\D]
\EE
which agrees with the restriction of (\ref{3}) to homogeneous variables
using $\B p^{ab}=\int_Up^{ab}$. Since the time derivatives of $A$ and $N$ do
not appear in the Lagrangian one finds the primary constraints
\BE
       p_{A^\mu}\approx 0\qquad\qquad p_N\approx 0
\EE
and the associated secondary constraints
\BEA
      {C_D}_\mu&=&\{H,p_{A^\mu}\}=-2\Tr qr_\mu\B p\approx 0\\
      C_H&=&\{H,p_N\}=\F{1}{V}\G(\Tr (\B pq)^2-{\TS\F{1}{2}}\Tr\!^2\B
      pq\D)-VR\approx 0\qquad.
\EEA
Thus the Hamiltonian is, as usual, a sum of the constraints.
\BE
    H(q,\B p,A,N)=-2A^\mu{C_D}_\mu+NC_H=:H_D(A)+H_H(N)\approx 0
\EE
The diffeomorphism part in the Hamiltonian directly equals the momentum
mapping of the action of Aut\BBB\ on the configuration space, if one
identifies $\pi$ and $\B p$. An important property of the momentum mapping
can be checked easily:
\BE
   \{H_D(A),H_D(B)\}=H_D([A,B])
\EE
Thus the constraints ${C_D}_\mu$ commute weakly:
\BE
   \{{C_D}_\mu,{C_D}_\nu\}=\alpha^\rho{}_{\mu\nu}{C_D}_\rho\qquad\qquad
     [r_\mu,r_\nu]=:\alpha^\rho{}_{\mu\nu}r_\rho
\EE
The identification of $\pi$ and $\B p$ leads to another difficulty: $p$ is
a tensor valued 3-form, $\B p$ a tensor valued volume, and so it transforms
differently. This is the reason why the diffeomorphism constraint does not
generate the Lie derivative of $\B p$, as one expects according to
ADM-theory.
\BD
   \B p\in{T^2_e}_{\SSS sym}G\otimes \Lambda^3(T_eG)=:W \qquad
   \Phi^W:\Autg\times W \lra W
\ED
\vspace{-25pt}
\BEA
    \Phi(S,\B p)&=&\det S^{-1}S\B pS^T\\
    \F{d}{dt}\St{t=0}\Phi(e^{tA},\B p)&=&-\Tr A\B p+A\B p+\B pA^T=L_\V N\B p
\EEA
This would also be the restriction of the diffeomorphism part of the
ADM-equations of motion to homogeneous variables. The first term is not
generated by $H_D(A)=-2\Tr qA\B p$, but it vanishes if one only considers the
unimodular subgroup of SAut\BBB\ of Aut\BBB. The restriction to the
unimodular subgroup is a natural consequence of a tertiary constraint which
has no equivalent in ADM-theory:
\BEAS
   \{H,C_H\}&=&A^\mu\{{C_D}_\mu,C_H\}=
     -A^\mu\F{d}{dt}\St{t=0}C_H\circ\Phi^{T^*}(\exp tr_\mu,(q,\B p))\\
     &=&-A^0 \Tr r_0\G({\TS\F{1}{V}}\G(\Tr (\B pq)^2
      -{\TS\F{1}{2}}\Tr^2(\B pq)\D)+VR\D)\approx 0\\
     &&\Lra\qquad A^0\approx 0
\EEAS
Here we used the fact that ${C_D}_\mu$ acts as a momentum mapping for the
action of Aut\BBB, where $\B p$ is treated as a symmetric tensor not as a
tensor valued volume. The group operation leaves the traces and the scalar
$R$ invariant. So $A^0\approx 0$ is a tertiary constraint and of course the
pair $(A^0,p_{A^0})$ is second-class. Hence one can discard this degree of
freedom and restrict to the unimodular subgroup of the automorphism group.

Finally we derive the equations of motion and compare them with the
restriction of the ADM-equations to homogeneous variables.
\BEA
\NN \dot{q}_{ab}=~\F{\P H}{\P \B p^{ab}}&=&
     -2q_{c(a}A^c{}_{b)}+\F{2N}{V}\G(\B p_{ab}-{\TS\F{1}{2}}\B p q_{ab}\D)\\
\NN   \dot{\B p}^{ab}=-\F{\P H}{\P q_{ab}}&=&
     +2A^{(a}{}_c\B p^{b)c}
     -\F{2N}{V}\G(\B p^a{}_c\B p^{cb}-{\TS\F{1}{2}}\B p^{ab}\B p\D)
  +\F{N}{2V}\G(\B p^{cd}\B p_{cd}-{\TS\F{1}{2}}\B p^2\D)q^{ab}\\
\label{4}
  &&\hspace{5ex}-NV\G(R^{ab}-{\TS\F{1}{2}}Rq^{ab}\D)
        +2NV\G(2a^aa^b-a^cC^{(a}{}_c{}^{b)}\D) \qquad
\EEA
The ADM-equations of motion are
\BEA
\NN  \dot{q}_{ab}=\F{\delta H}{\delta p^{ab}}&=&L_\V Nq_{ab}+2N\G(*p_{ab}
              -{\TS\F{1}{2}}pq_{ab}\D)\\
\NN  \dot{p}_{ab}=-\F{\delta H}{\delta q_{ab}}&=&L_\V Np^{ab}
               -2N\G(*p^{ac}p_c{}^b-{\TS\frac{1}{2}}pp^{ab}\D)
        +\F{N}{2}\eta q^{ab}\G(\GA p^{cd}|p_{cd}\DA-{\TS\frac{1}{2}}p^2\D)\\
     && -N\eta\G(R^{ab}-{\TS\frac{1}{2}}Rq^{ab}\D)
      +\G(\nabla^a\nabla^bN-q^{ab}\nabla^c\nabla\!_cN\D)\eta~~~~~~.
\EEA
If $\V N\in\mbox{saut}G,~N=N(t)$ and one integrates the second equation over
$U\subset G$ the equations differ only in the term
$2NV(2a^aa^b-a^cC^{(a~b)}_{~~c})$, which vanishes for class A algebras. This
term originates from the variation of the Ricci-tensor:
\BE
   N\delta\! R_{ab}q^{ab}
   =N(q^{ac}q^{bd}-q^{ab}q^{cd})\nabla\!_a\nabla\!_b\delta q_{cd}
\EE
Usually integration by parts leads to the term
$(\nabla^a\nabla^bN-q^{ab}\nabla^c\nabla\!_cN)\eta$, but for left invariant
variations one can evaluate $\nabla\!_a\nabla\!_b\delta q_{cd}$ using
(\ref{2}) for the Christoffel symbols. This leads to:
\BE
 N\delta\! R_{ab}q^{ab}=N\delta q_{ab}\G(4a^a a^b-2a^cC^{(a}{}_c{}^{b)}\D)
     =:-NQ^{ab}\delta q_{ab}
\EE
In the ADM-theory the equation for $\dot{p}^{ab}$ is equivalent to Einstein's
equation $G(b_a,b_b)\linebreak =0$. Thus equation (\ref{4}) is usually
corrected \cite{JII} by a term that cancels the additional contributions
derived from a left invariant action principle:
\BEAS
    \dot{q}_{ab}&=&\{q_{ab},H\}\\
    \dot{\B p}^{ab}&=&\{\B p^{ab}, H\}+NVQ^{ab}-\B p^{ab}\Tr A \qquad
    \Longleftrightarrow \qquad
    G^{ab}=0\qquad\not\hspace{-6pt}\Longleftrightarrow
    (\ref{4})
\EEAS
As we have shown the correction by the second term would not be necessary,
since $\Tr A$ vanishes in the physical part of the phase space. Since
$Q^{ab}dq_{ab}$ -- the exterior derivative refers to the configuration space
$Q$ where $q_{ab}$ are coordinates -- is not exact, one cannot substitute
$H$ by $H'=H+U$, s.t. $\{\B p^{ab},H\}+NV=\{\B p^{ab},H'\}$. So $Q^{ab}$
plays the role of a non-conservative force in this approach \cite{Sneddon}.
But one could also suppose that Einstein's equation $G(b_a,b_b)=0$
does not hold for class B algebras and accept equation (\ref{4}). As we have
seen the ADM-diffeomorphism-constraint equivalent to $G(n,b_a)=0$ does not
generate automorphisms in case of class B algebras. If one regards Einstein's
equations as a consequence of the action principle one could accept to reject
the Einstein-equations in the homogeneous case and to work with the
equivalents we have given. One should also take into account that the
assumptions of the model are non-covariant, since covariance and the
existence of a natural foliation are incompatible. It is the question what is
meant by reducing Einstein's theory to homogeneous cosmologies: Reducing
Einstein's equations or reducing the action principle from which it can be
derived.


\end{document}